\documentclass[twocolumn,showpacs,amsmath,amssymb]{revtex4}

\usepackage{graphicx}
\usepackage{dcolumn}
\usepackage{bm}
\newcommand{\etal}{\emph{et al.}}
\newcommand{\be}{\begin{equation}}
\newcommand{\ee}{\end{equation}}
\newcommand{\bfig}{\begin{figure}}
\newcommand{\efig}{\end{figure}}
\newcommand{\incl}{\includegraphics}

\begin{document}      

\title{The zero-energy state in graphene in a high magnetic field}

\author{Joseph G. Checkelsky, Lu Li and N. P. Ong
}
\affiliation{
Department of Physics, Princeton University, Princeton, NJ 08544, USA
}
\date{\today}      
\begin{abstract}
The fate of the charge-neutral Dirac point in graphene in a high 
magnetic field $H$ has been investigated at low temperatures ($T\sim$ 0.3 K).  
In samples with small gate-voltage offset $V_0$, the resistance $R_0$ at the Dirac point diverges steeply with $H$, signalling a 
crossover to an insulating state in high field.  
The approach to the insulating state is highly unusual.  
Despite the steep divergence in $R_0$, the profile of $R_0$ vs. $T$ in
fixed $H$ saturates to a $T$-independent value below 2 K, consistent 
with gapless charge-carrying excitations.
\end{abstract}

\maketitle                   
The discovery of the quantum Hall effect (QHE) in 
monolayer graphene crystals provides a new system for investigating
relativistic Dirac-like excitations in solids
\cite{Novoselov1,Novoselov2,Novoselov3,Zhang1,Zhang2,Tan}. 
In a magnetic field $H$, the system forms
Landau Levels (indexed by $n$) that are 4-fold degenerate.
The Hall conductivity $\sigma_{xy}$ is accurately quantized 
as the chemical potential $\mu$ is changed from the hole part
to electron part of the Dirac spectrum.  Considerable attention
has focussed on the $n=0$ Landau Level (LL), especially on the nature
of the electronic state at the charge-neutral point ($\mu = 0$) in an intense magnetic field $H$.
Several groups~\cite{MacDonald,Fisher,Yang,Goerbig,AbaninLee,YangRev} 
have predicted that a large field stabilizes the 
quantum Hall ferromagnetic (QHF) state, in which the pseudospins 
describing the valley degree of freedom become 
ferromagnetically ordered. In a second group of theories~\cite{Khveshchenko,Miransky}, interaction
causes an excitonic gap to open at the Dirac point.  
Experiments are actively addressing these issues~\cite{AbaninGeim,Jiang,Fuhrer}.  Jiang \etal~\cite{Jiang} have inferred that the sublevel gaps at $\nu$ = 0 and $\pm$1 arise
from lifting of the spin and sub-lattice degeneracies, respectively, and inferred a many-body origin for the states.  
We have found that, in samples with small $V_0$ (the 
gate voltage needed to align $\mu$ with the 
Dirac point), the value $R_0$ of the resistance $R_{xx}$ at the 
Dirac point diverges steeply with $H$, i.e. a large $H$ drives the
Dirac point insulating.  Despite the strong $H$ dependence, $R_0$ saturates to a $T$-independent value below 2 K, providing evidence for charged, gapless excitations. In samples with large $V_0$, this divergence in $R_0$ 
is shifted to higher fields.

\bfig[h]			
\incl[width=8cm]{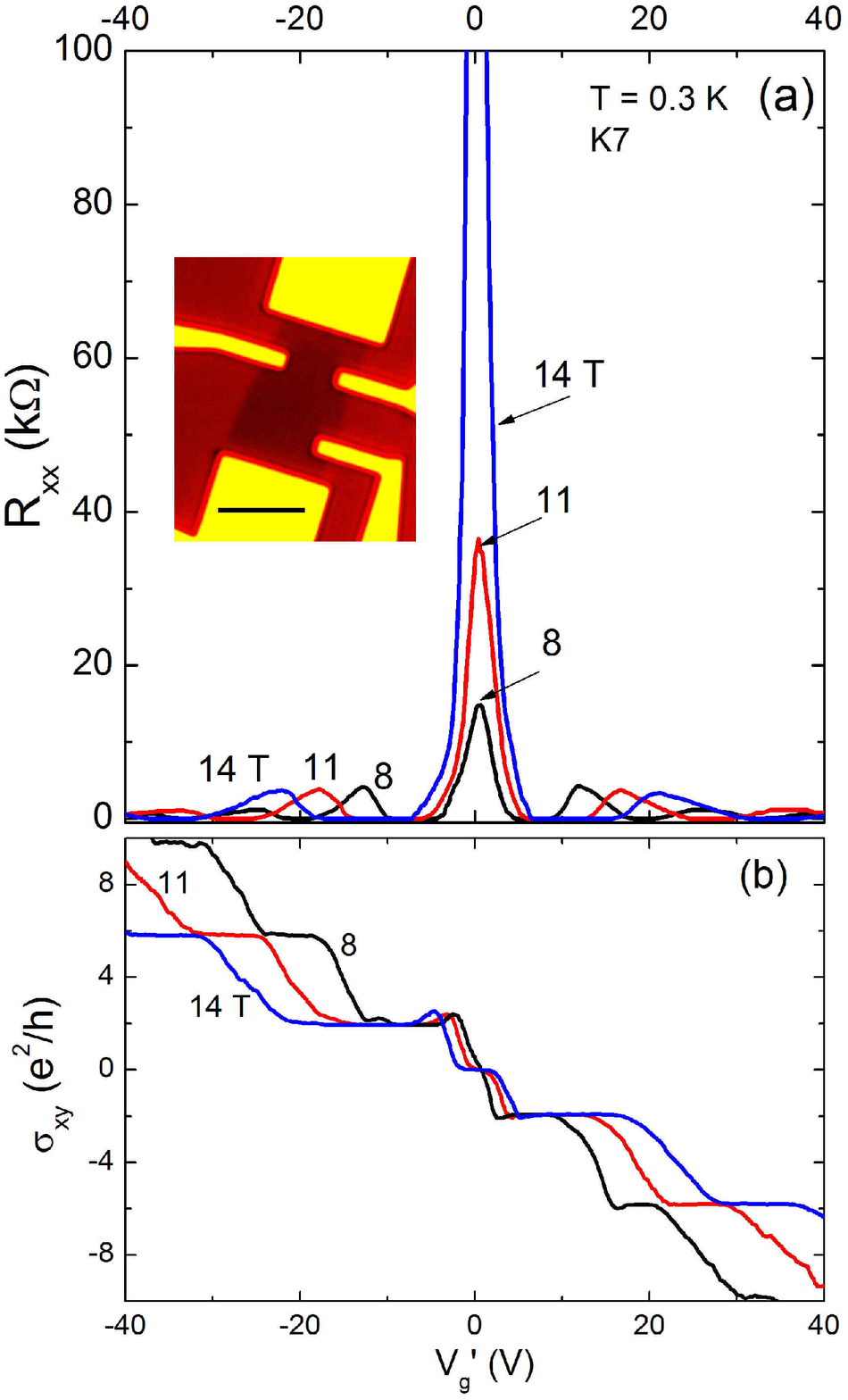}
\caption{\label{figRS} (color online)
The resistance $R_{xx}$ (Panel a) and Hall conductivity
$\sigma_{xy}$ (b) in Sample K7 versus (shifted) gate voltage $V_g'= V_g-V_0$
at 0.3 K with $H$ fixed at 8, 11 and 14 T.  Peaks of $R_{xx}$ at finite
$V_g'$ correspond to the filling of the $n=1$ and $n=2$ LLs.  
At $V_g' = 0$, the peak in $R_{xx}$ grows to 190 k$\Omega$ at 14 T.
The inset shows in false color a graphene crystal (dark red) with Au leads 
deposited (yellow regions). The bar indicates 5 $\mu$m. 
Panel b shows the quantization of $\sigma_{xy}$ at the values
$(4e^2/h)(n+\frac12)$.  At 0.3 K, $\sigma_{xy} = 0$ in a
a 2-Volt interval around $V_g' = 0$. 
}
\efig
%
%

Following Refs.~\cite{Novoselov1,Novoselov2,Zhang1}, 
we peeled single-layer graphene crystals
(3-10 $\mu$m in length) from Kish graphite on a Si-SiO$_2$ wafer.  
Au/Cr contacts were deposited using e-beam lithography (Fig. \ref{figRS}a, inset).  
We have found that the high-field 
behavior of $R_0$ is strongly correlated with 
$V_0$ (Table \ref{tab}).  All samples (except K22) 
have $\mu$ lying in the electron band (positive $V_0$).  
Samples in which $|V_0|<$ 1 V (K7 and K22)
display a very large $R_0(14)$ (resistance measured at 14 T and
0.3 K), which arises from the strong divergence mentioned.
By contrast, in samples with large $|V_0|$, $R_0(14)\le$ 7 k$\Omega$.

\begin{table}[b]
\begin{tabular}{|c|c|c|c|}		\hline
Sample	& $V_0$(V)& $R_0(14)$(k$\Omega$)& $\mu_e$(1/T) \\ \hline

K5			& 3				& 80			& 0.3		\\
K7			&	1				&	190			&	1.3    \\		
K8			& 12			& 15			&	0.6    \\ 
K18			& 20			& 7.5			&	0.9    \\ 
K22			& -0.6			& $>$280	&	2.5    \\ 
K29			& 22.5			& 7			&	0.2    \\ 
\hline
\end{tabular}
\caption{\label{tab}
Sample parameters.  $V_0$ is the gate voltage needed to bring $\mu$ to 0.  
$R_0(14)$ is $R_0$ measured at $H$ = 14 T and $T$ = 0.3 K. 
$\mu_e$ is the electron mobility at $H=0$.
}
\end{table}
%

\bfig[h]			
\incl[width=8.5cm]{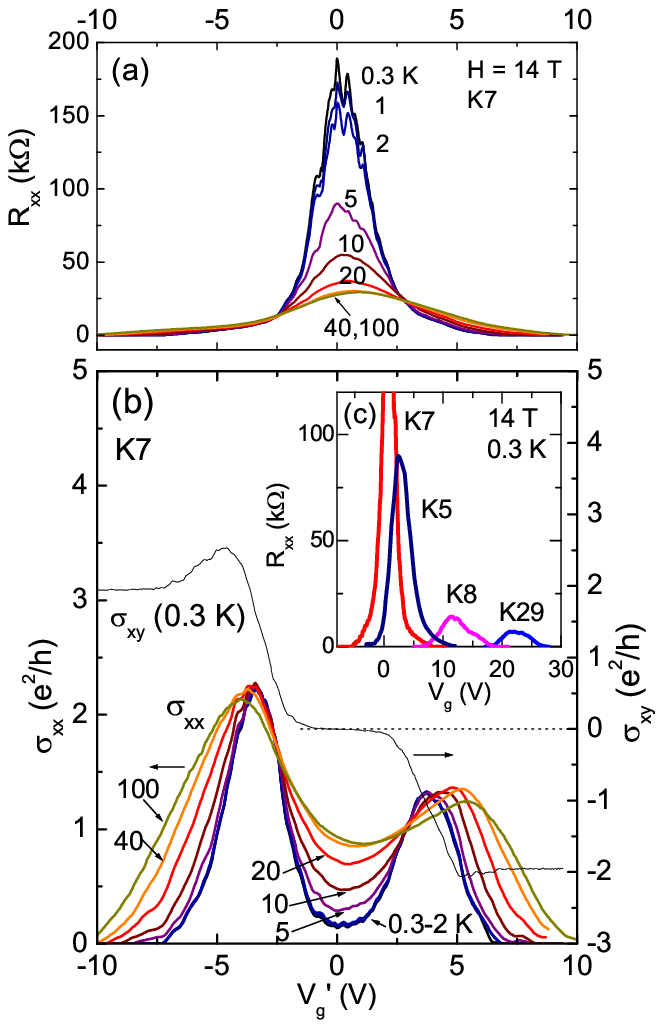}
\caption{\label{figR0} (color online)
The resistance $R_{xx}$, conductivity $\sigma_{xx}$
and the Hall conductivity $\sigma_{xy}$ in K7 vs. the shifted
gate voltage $V_g'$, with $H$ fixed at 14 T.  
As $T$ decreases to 0.3 K, the
zero-energy peak in $R_{xx}$ (Panel a) rises steeply to 190 k$\Omega$.
Panel (b) shows that, as $T$ decreases, double peaks in $\sigma_{xx}$ are
clearly resolved. Between the peaks, $\sigma_{xx}$ falls rapidly but 
saturates below 2 K.  The Hall conductivity at 0.3 K (thin curve) 
displays a clear plateau ($|\sigma_{xy}|<0.02 e^2/h$) in the 
interval $-1 V <V_g'<1 V$.  
Panel (c) compares $R_{xx}$ (of $n=0$ LL) vs. unshifted gate $V_g$ in 
the samples K5, K7, K8 and K29 at 0.3 K. In each sample, $R_{xx}$ peaks
at $V_0$.  As $V_0$ increases, $R_0(14)$ rapidly decreases.
}
\efig
%
%

Figure \ref{figRS}a shows the variation of $R_{xx}$ in K7 plotted vs. 
the shifted gate voltage $V_g'=V_g-V_0$ 
with $H$ held at 8, 11 and 14 T (at $T$ = 0.3 K).  
The striking feature here is that the peak corresponding to
the $n$ = 0 LL increases to $>$100 k$\Omega$
at 14 T, whereas the peaks corresponding to $n = \pm 1$ 
remain below $\sim$7 k$\Omega$.
As in Refs. \cite{Novoselov1,Novoselov2,Novoselov3,Zhang1,Zhang2}, the Hall conductivity $\sigma_{xy}$ (Panel b) displays plateaus given by~\cite{YangRev}
\be
\sigma_{xy} = \frac{\nu e^2}{h} = \frac{4e^2}{h}\left(n+\frac12 \right),
\label{sigma}
\ee
where $n$ indexes the 4-fold degenerate LL and $\nu$ indexes individual sublevels.  In K7, the `zero' plateau $\sigma_{xy} \simeq 0$ at $V_g'=0$ is already visible at $H =$ 8 T.

Narrowing our focus to the $n$ = 0 LL, we examine 
$R_{xx}$ in the $n=0$ LL as a function of $T$, with $H$ fixed at 14 T (Fig. \ref{figR0}a).  We see that, from 40 to 0.3 K, $R_0$ rises steeply from 4 k$\Omega$ to 190 k$\Omega$.
The curve of the conductivity $\sigma_{xx}$ plotted vs. $V_g'$ reveals a
2-peak structure that implies splitting of the 4-fold degeneracy by a gap
$\Delta$ (Fig. \ref{figR0}b).  At 100 K, the two peaks are already resolved.  With decreasing $T$, the minimum at $V_g'=0$ initially deepens rapidly, but 
saturates below $\sim$2 K.  The Hall conductivity $\sigma_{xy}$ at 0.3 K (thin curve)
displays a well-defined plateau on which $\sigma_{xy} \simeq 0$.
Because the next plateau is at $\sigma_{xy} = 2(e^2/h)$, we infer
that each of the peaks in $\sigma_{xx}$ is comprised of 2 unresolved sublevels. The opening of the gap causes $\sigma_{xx}$ (at $V_g' =0$) 
to fall rapidly with decreasing $T$, until saturation occurs below 2 K.  

The behavior of $R_0$ described here is qualitatively
different from that in, for e.g., Ref.~\cite{AbaninGeim}.  
To understand the difference, we have examined several samples 
(Table \ref{tab}).  As mentioned, the offset gate $V_0$ is a crucial
parameter.  Figure \ref{figR0}c compares the curves of $R_{xx}$ ($n=0$ LL) 
in the samples K5, K7, K8 and K29, all measured at 14 T at $T$ = 0.3 K.
For each sample, we have plotted $R_{xx}$ vs. the unshifted gate voltage $V_g$, 
so its peak automatically locates $V_0$.  
It is clear that K7 ($V_0$ = 1 V) has the highest peak, followed
by K5 ($V_0$ = 3 V), whereas K8 ($V_0$ = 12 V) and K29 (22.5 V) have
peaks that are severely suppressed.  Further insight into this 
pattern of suppression is given below when we examine 
the $H$ dependence of $R_0$ at low $T$.

\bfig[h]			
\incl[width=8cm]{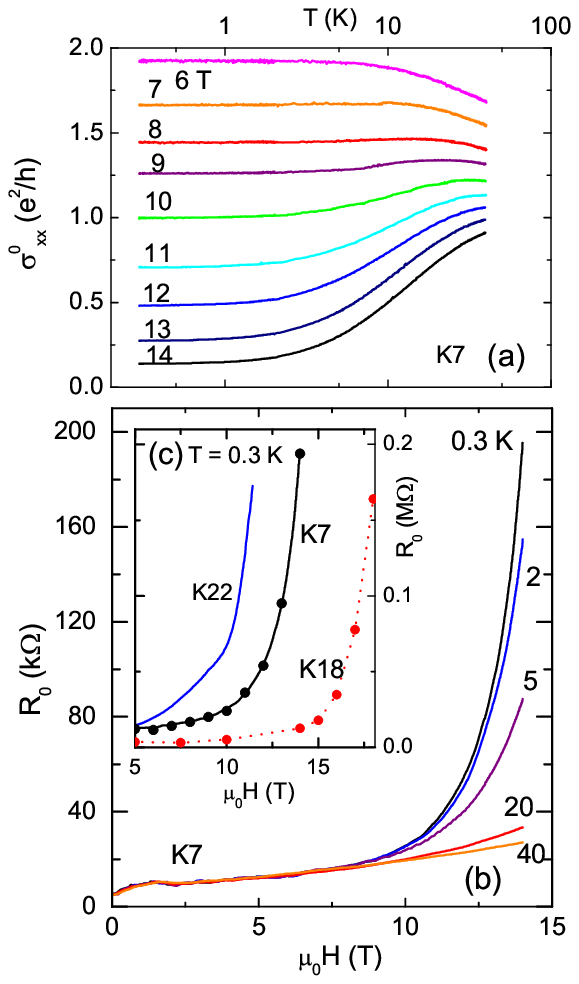}
\caption{\label{figG0} (color online)
The $T$ dependence of $\sigma_{xx}^0$ in K7
(= $L/wR_0$) and the $H$ dependence of $R_0$ at low temperature.
Panel a shows curves of $\sigma_{xx}^0$ vs. $\log_{10} T$  
with $H$ fixed at 6-14 T. For $H>$ 8 T, the gap $\Delta$ causes
$\sigma_{xx}^0$ to decrease markedly until saturation at the
residual value $\sigma_{res}$ occurs below 2 K.  
Panel (b) displays the steep increase in $R_0$ vs. $H$ in K7 
at selected $T$.  At 0.3 K, $R_0$ appears
to diverge at a field near 18 T (see Fig. \ref{contour}b).
Panel (c) compares the $R_0(H)$ profiles in Samples K7, K18 and K22.  
In Sample K18 ($V_0$ = 20 V), the divergence 
in $R_0$ becomes apparent only above 14 T, whereas in K22 ($V_0$ = -0.6 V)
$R_0$ starts to diverge at fields lower than in K7.  In K7, 
we have plotted $R_0$ values measured by sweeping $V_g$ at fixed $H$ 
(solid symbols) with $R_0$ measured by sweeping
$H$ with $V_g'$ fixed at 0 (solid curve), to show consistency.
}
\efig

\bfig[h]			
\incl[width=9cm]{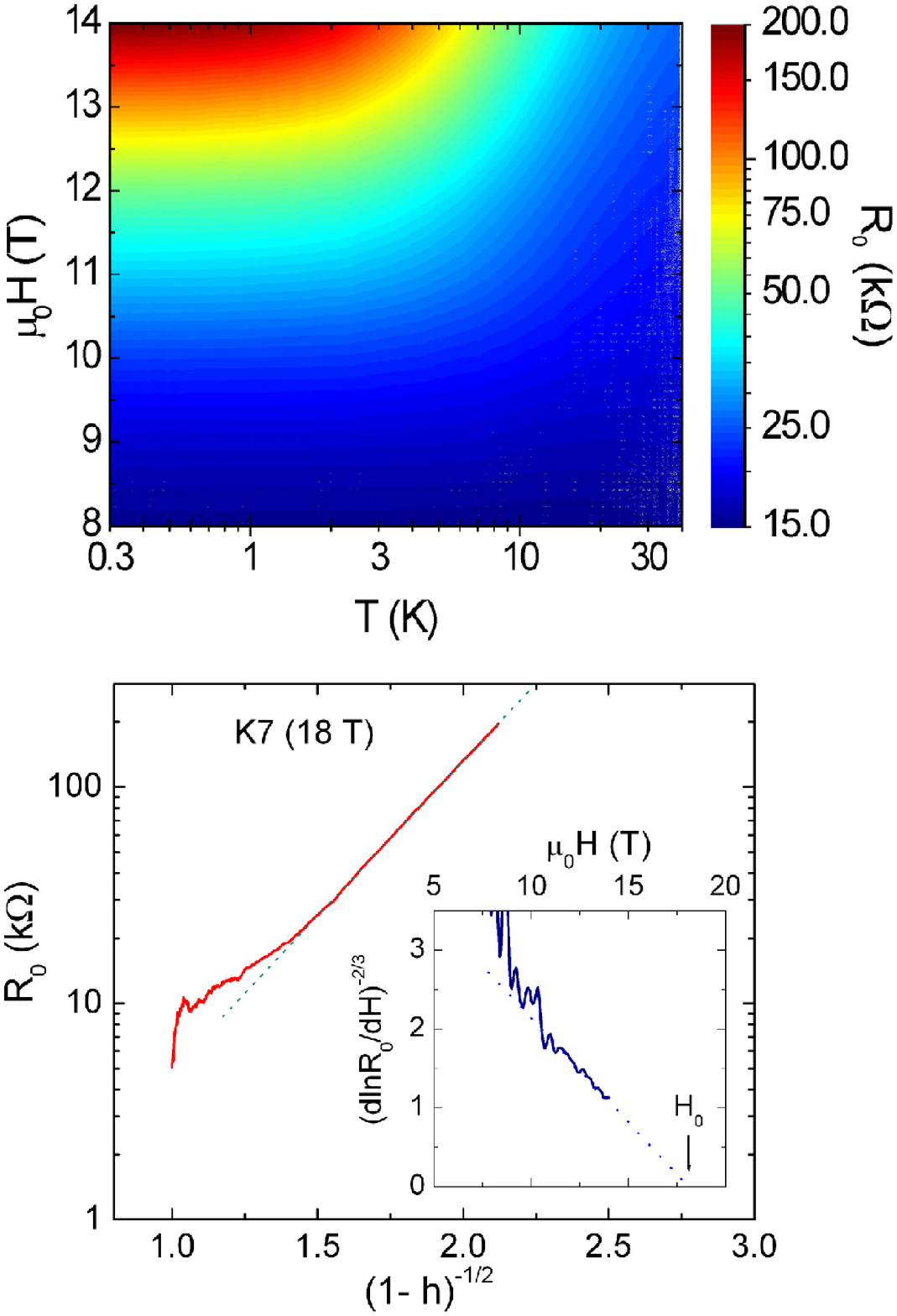}
\caption{\label{contour}  (color online)
(Panel a) The contour plot of $R_0(T,H)$ (K7) in the
$T$-$H$ plane (vertical bar shows values of $R_0$).  The contour lines
emphasize the unusual approach to the insulating state.  
At low $T$, $R_0$ is unchanged on a horizontal path 
($H$ held fixed), but it rises rapidly on a 
vertical path (increasing $H$ at fixed $T$). 
Panel (b) displays $\log R_0$ vs. 1/$\sqrt{1-h}$ in K7
at $T$ = 0.3 K, with $h = H/H_0$, where 
$H_0$ = 18 T.  The linear segment at large $R_0$ shows that
the divergence is consistent with $R_0(h)\sim \exp[{2b/\sqrt{1-h}}]$ with $b\sim$0.7.  In Panel (c), the plot of $(d\ln R_0/dH)^{-2/3}$ vs. $H$
shows a high-field linear segment that 
extrapolates to zero at $\sim H_0$ (18 T).
}
\efig

Sample self-heating may also obscure the divergence.  We find that,
below 1 K, self-heating becomes serious when the dissipation exceeds $\sim$2 pW.
The measurements of $R_{xx}$ vs. $V_g$ were repeated 
at 3 currents ($I$= 0.6, 2 and 15 nA)
at $T$ = 0.3 K.  The results at $I$ = 0.6 and 2 nA are virtually identical. 
However, the curve at 15 nA is 30$\%$ smaller near $V_g' = 0$, consistent
with heating.  Hence, we have kept $I$ at 2 nA to eliminate 
self-heating as a problem.  Heating at the contacts is negligible because
of the small contact resistances ($\sim$1 k$\Omega$) relative to $R_0$.

Hereafter, we focus on $R_0$, or equivalently, the Dirac-point conductivity
$\sigma_{xx}^0 \equiv L/wR_0$ ($L$ and $w$ are the length
and width).  Curves of the conductivity versus $\log_{10}T$ are 
shown in Fig. \ref{figG0}a at selected fields.  In low fields ($H<$ 9 T),
the $T$ dependence of $\sigma_{xx}^0$ is quite mild.  As 
$H$ is increased to 14 T, the opening of the gap $\Delta$ (between the
$n$=0 sublevels) causes the conductance to decrease sharply below 40 K. 
However, instead of falling to 0, $\sigma_{xx}^0$ saturates below 2 K to a $T$-independent 
residual value $\sigma_{res}$, as anticipated in the discussion of
Fig. \ref{figR0}b.  The existence of this
residual $\sigma_{res}$, which is highly sensitive to $H$, is 
one of our key findings.  

The field dependence of $\sigma_{res}$ is best 
viewed as a divergent $R_0$.  Figure \ref{figG0}b shows the rising
profile of $R_0$ vs. $H$ in sample K7 at selected temperatures.  
The divergent form at the lowest $T$ (0.3 K) strongly suggests that the system 
is rapidly approaching a field-induced crossover (or transition) 
to an insulating state.  

In light of the importance of $V_0$, it is instructive to see how
the profile of $R_0$ vs. $H$ varies between samples.
Figure \ref{figG0}c compares the results in K7, K18 and K22 at $T$ = 0.3 K.
In K18, where $V_0$ (20 V) is quite large, the divergence 
in $R_0(H)$ becomes noticeable only in fields above 14 T.
Conversely, in K22 for which $V_0$ (-0.6 V) is slightly smaller than in K7, 
$R_0$ diverges at field scales smaller than in K7.  
From the trend, it is clear that the divergence
in $R_0$ is shifted to ever higher fields as $V_0$ increases.  Referring
back to Fig. \ref{figR0}c, we now see that the strong suppression of $R_0(14)$
in samples with large $V_0$ simply reflects the shift of the divergence to larger $H$.  These results underscore the importance of choosing samples with $|V_0|<$ 1 V for investigating the intrinsic properties of the Dirac-point.
In the case of K7, we have also checked that the 2 procedures
for measuring $R_0$ (varying $V_g$ at fixed $H$ or vice versa)
give consistent values.   

The variation of $R_0(T,H)$ in K7 is conveniently represented in a contour plot 
in the $T$-$H$ plane (Fig. \ref{contour}). Below $\sim$2 K, the contour lines are 
horizontal, which implies that $R_0$ is unchanged if the sample is cooled in fixed $H$.
This provides evidence that $\sigma_{res}$ involves gapless excitations.  
However, if $T$ is fixed, $R_0$ rises steeply with $H$, implying proximity
to the insulating state (deep-red region). When a system approaches the insulating state,
its resistivity generally diverges as $T\rightarrow 0$, as a result of either
strong localization (variable-range hopping) or
the opening of a mobility gap (weak localization is not relevant here
because of the intense $H$).  In both cases, decreasing $T$ reduces
the conductance because the itinerant states are severely depopulated.
Hence, the pattern in Fig. \ref{contour}a is most unusual.  The gaplessness of $\sigma_{res}$ suggests that, below 2 K, these excitations are protected from the effects of changing $T$.  Paradoxically, they are not protected
from an increasing $H$, which reduces the current carried at an exponential rate.

In the theory in Refs.~\cite{Abanin,AbaninGeim}, the current at the Dirac point is 
carried by a pair of edge states.  The change in $R_{xx}$ vs. $H$ is 
interpreted as an increased scattering rate
as the edge states are pushed closer to the edge.

In samples with small $V_0$, however, the steep increase 
in $R_0$ to $\sim$200 k$\Omega$ (Fig. \ref{figG0}b) 
suggests a different regime in which Coulomb exchange 
may be the dominant energy scale.  
From Figs. \ref{figG0}b and \ref{contour}a, we infer that $R_0$
appears to be diverging towards an insulating state in high fields.
These considerations lead us to quantify the divergence in $R_0$.

We find that $R_0$ fits very well to the form $R_0\sim \xi(h)^2$, 
where the correlation length $\xi$ has the Kosterlitz-Thouless (KT) form
\be
\xi_{KT} \sim \exp(b/\sqrt{1-h}), \quad (h = H/H_0),
\label{KT}
\ee
with $H$ replacing $T$.  Plotting $\ln R_0$ vs. $\sqrt{1-h}$, 
we find that the high-field portion becomes linear (Fig. \ref{contour}b)
when $H_0$ is adjusted to be 17-18 T. From the slope, we find that
the parameter $b$ has the value $\sim$0.7, in agreement with simulations of
the KT transition.  For self consistency, 
we may also let the data determine $H_0$ by linear extrapolation.  
By Eq. \ref{KT}, we have $d\ln R_0/dH\sim (H_0-H)^{-3/2}$. 
Hence a plot of $(d\ln R_0/dH)^{-2/3}$ vs. $H$ should cross the $H$-axis
at $H_0$.  Indeed, this quantity, plotted in Fig. \ref{contour}c,
becomes linear at large $H$ and extrapolates to zero at $\sim$18 T 
in agreement with (b).  

Although consistency with Eq. \ref{KT} alone does not prove 
that a KT transition occurs at $H_0$, the fit does reveal the striking exponential
character of the divergence in $R_0$.  We adopt as a working hypothesis that this reflects the approach to a KT transition.  The ordered state 
at large $H$ is destroyed by the spontaneous appearance of defects 
which increase exponentially in density at $H_0$.
The steep fall of $R_0$ below $H_0$ may reflect the current carried by these defects.

In both the QHF state~\cite{MacDonald,Fisher,Yang,Goerbig,AbaninLee} 
and the exciton-gap state~\cite{Khveshchenko,Miransky},
the 4-fold degeneracy in the $n=0$ LL is completely 
lifted in large $H$ to produce an insulator.  In the QHF, the starting 
symmetry is $SU(4)$ if we treat the spin and pseudospin degrees on equal footing~\cite{Yang}.
The reduction of the $SU(4)$ to lower symmetries by the effects of
Zeeman energy, disorder or lattice discretization is discussed in Refs.~\cite{Goerbig,Fisher,AbaninLee,YangRev}.  According to Ref. \cite{AbaninLee}
random disorder may force the pseudospins into the plane.
If the ordered state indeed has $U(1)$ symmetry, the XY order is susceptible
to a KT transition~\cite{MacDonald,AbaninLee}.  It would be very interesting
to relate the fit of $R_0$ to Eq. \ref{KT} with charged, topological
excitations envisioned in the KT transition.
Measurements are in progress at higher $H$ to
provide further evidence for a transition to the insulating state
and to clarify its nature.

We acknowledge valuable discussions with Kun Yang, P. A. Lee, M. Lee, 
A. Geim and A. Pasupathy.  We are indebted to Philip Kim 
for generously providing Kish graphite crystals.  This research 
is supported by NSF-MRSEC under Grant DMR 0213706,
and by the Princeton Center for Complex Materials.  Some data
were taken in the National High Magnetic Field
Lab., Tallahassee, which is supported by NSF and the State of Florida.
%
%
%

%

%
\end{document}